\documentclass{osa-article}

\journal{oe}

\usepackage{ulem}
\usepackage{footnote}

\begin{document}

\title{Broadband high-resolution terahertz single-pixel imaging}

\author{Adam Vall\'es,\authormark{1,*} Jiahuan He,\authormark{2} Seigo Ohno,\authormark{3} Takashige~Omatsu\authormark{1,2} and Katsuhiko Miyamoto,\authormark{1,2,$\dagger$}}

\address{\authormark{1}Molecular Chirality Research Center, Chiba University, 1-33 Yayoi-cho, Inage-ku, Chiba 263-8522, Japan\\
\authormark{2}Graduate School of Advanced Integration Science, Chiba University, 1-33 Yayoi-cho, Inage-ku, Chiba 263-8522, Japan\\
\authormark{3}Graduate School of Science, Tohoku University, Sendai 980-8578, Japan}

\email{\authormark{*}adam.valles@chiba-u.jp}
\email{\authormark{$\dagger$}k-miyamoto@faculty.chiba-u.jp}



\begin{abstract}
We report a simple single-pixel imaging system with a low mean squared error in the entire terahertz frequency region (3-13 THz) that employs a thin metallic ring with a series of directly perforated random masks and a subpixel mask digitization technique. This imaging system produces high pixel resolution reconstructed images, up to 1200 $\times$ 1200 pixels, and imaging area of 32 $\times$ 32 mm$^2$. It can be extended to develop advanced imaging systems in the near-ultraviolet to terahertz region.
\end{abstract}

\section{Introduction}
There has been increasing demand for imaging systems in the terahertz (THz) region, in which a variety of molecules and their aggregations exhibit stretching and vibration eigen frequencies \cite{chan2007imaging}. Terahertz imaging systems allow the non-invasive identification of materials and structures in biological tissues, and thus facilitate the development of two-dimensional (2D) scanners for pathological diagnosis and illegal drug detection systems \cite{loffler2001terahertz, pickwell2004vivo, kawase2003non}. The aforementioned applications require a terahertz imaging system with a high signal-to-noise ratio (SNR) or a low mean squared error (MSE) \cite{candes2006compressive} in the entire terahertz frequency region. 2D terahertz imaging systems based on microbolometer arrays, called THz cameras \cite{oda2015microbolometer}, have been developed, however, it is difficult to capture high pixel resolution images, for low power signals, in the entire high-frequency region above 4 THz.

Although THz cameras are the imaging technology most commonly used, they require high signal intensity to overcome the inherent noise in each microbolometers, i.e., $\sim$ 20 pW minimum detectable power per pixel at 4 THz. Such sensitivity could notably be improved by using a cryogenic Si bolometric detector \cite{downey1984monolithic}, being able to consider $\sim 14$ fW (or $\sim 5$ pW) minimum detectable power within most of the THz region if cooling the detection system down to 0.3 K (or 4.2 K). However, this type of THz detectors are bulky and cannot be implemented in a 2D array.

Single-pixel imaging (SPI) is an imaging technique for reconstructing compressive images that employs a series of spatially resolved patterns to mask the object of interest while measuring the overall transmitted signal with a bucket detector, which lacks spatial resolution \cite{duarte2008single}. Single-pixel cameras \cite{duarte2008single, edgar2019principles} allow the development of highly sensitive detection systems \cite{donoho2006compressed, candes2006near} for applications such as hazardous material detection \cite{stantchev2016noninvasive, gibson2017real}, holography \cite{clemente2013compressive}, polarimetry \cite{soldevila2013single}, microscopy \cite{studer2012compressive,radwell2014single}, ghost imaging \cite{bornman2019ghost}, and three-dimensional video processing \cite{sun2016single}.

Several terahertz SPI systems have been proposed. Structured illumination based on optical-to-terahertz nonlinear conversion has been demonstrated, however, the use of a nonlinear crystal (ZnTe) strongly impacts imaging at frequencies of above 2.5 THz \cite{olivieri2020hyperspectral}. An optically controlled dynamic spatial light modulator also does not permit high-quality imaging at frequencies of above 2 THz \cite{shrekenhamer2013terahertz}. A spinning disk that includes random copper patterns printed on a standard printed circuit board (PCB) substrate for terahertz compressive image acquisition has been proposed. However, it allows the reconstruction of images only at frequencies of below 0.8 THz owing to the strong absorption loss of the PCB substrate \cite{chan2008single, shen2012spinning}. Thus, it is difficult to reconstruct images with high quality while keeping the spectral information.

Here, we propose a simple terahertz SPI scheme to realize high pixel resolution 2D imaging in the entire terahertz frequency region (3-13 THz), using a high-sensitivity single-point THz detector. A metallic ring with a series of directly perforated random masks, encoded with almost equally weighted ON:OFF pixels, is employed. We define the pixel resolution as the number of pixels used to represent a given imaging area; hence, an increase in pixel resolution enables us to distinguish smaller details in an image \cite{sun2016improving, phillips2017adaptive}. It is worth noting that this pixel resolution is different from the spatial resolution, which is determined by the inherent diffraction effects.

\section{Implementation}

\subsection{Concept and principle}

Despite the lack of spatial resolving capabilities in such high-sensitivity cryogenic Si bolometric detectors, other imaging methods can be considered to reconstruct any given object. The simplest one would be to perform a single-pixel raster scan (RS). This involves transversely displacing a transmitting window (surrounded by a non-transmitting material) while recording the transmitted signal for each transverse position \cite{hu1995imaging}. The problem is that the transmitting window should be displaced as many times as the number of pixels in the image ($N = n \times n$). Thus, depending on the desired pixel resolution, e.g. $N = 320 \times 320 \approx 10^5$ in our case, this method can be extremely time-consuming. Furthermore, a higher resolution leads to smaller pixels, i.e., a lower transmitted signal, making it difficult to detect the signal even with high-sensitivity detection systems \cite{petrov2016application}. However, this is not the case if considering a single-pixel imaging (SPI) system where the spatially resolved random patterns used to project the object is formed by 50:50 proportion of transmitting and blocking pixels (ON:OFF). Meaning that the detected power is 50\% of the total, on average, and is independent of the selected image resolution (pixel size). In other words, we can increase the pixel resolution of the final image as much as we want without affecting the signal-to-noise ratio (SNR), but at the cost of needing more measurements.

As in most SPI reconstruction methods, we aim to reduce the number of measurements required to obtain a high pixel resolution image when the signal intensity, which contains the spatial information, is low. Several algorithms have been proposed for reducing the total number of measurements, $M$, to less than the total number of pixels in an image, $N$ ($= n \times n$). Such optimization methods are beyond the scope of this manuscript (see the Appendix for more details). Thus, we simply designed a concatenation of independent and identically distributed [0,1] random matrices from an uniform Bernoulli distribution \cite{baraniuk2008simple}, used in the following object reconstruction algorithm:
\begin{equation}
  \mathbf{O} \approx \mathcal{T} \sum_i^M\frac{(x_i - \Bar{x})}{\xi_i} \mathbf{\mathbf{\Phi}}_i,
\label{eqn:randommaskimage}
\end{equation}
\noindent where $\mathcal{T}$ is the transmission of ON pixels that depends on the optical frequency, $x_i$ is the transmitted optical intensity passing through each random mask $\mathbf{\Phi}_i$ and object $\mathbf{O}$ (being both $\mathbf{\Phi}_i$ and $\mathbf{O}$, an $n \times n$ matrix), $\Bar{x}$ is the averaged $x_i$, and $\xi_i$ is a normalization coefficient related to the ratio of ON:OFF pixels in each mask. In the frequency range in which a supporting material exhibits large absorption, or under the cut-off frequency of a metal hole, $\mathcal{T}$ gives a low value and degrades image quality. The object $\mathbf{O}$ consists of a correlation of the terahertz intensity detected after traversing a given object and masked with a series of random patterns. A random mask with a spatial pattern similar to that of the object will thus be correlated to a higher transmitted intensity, and hence heavily weighted in the reconstruction algorithm. Note that the pixel resolution of the reconstructed image is directly related to the number of elements ($n \times n$) in the random masks $\mathbf{\Phi}$ used.



\begin{figure}[t!]
\centering
\includegraphics[width=\linewidth]{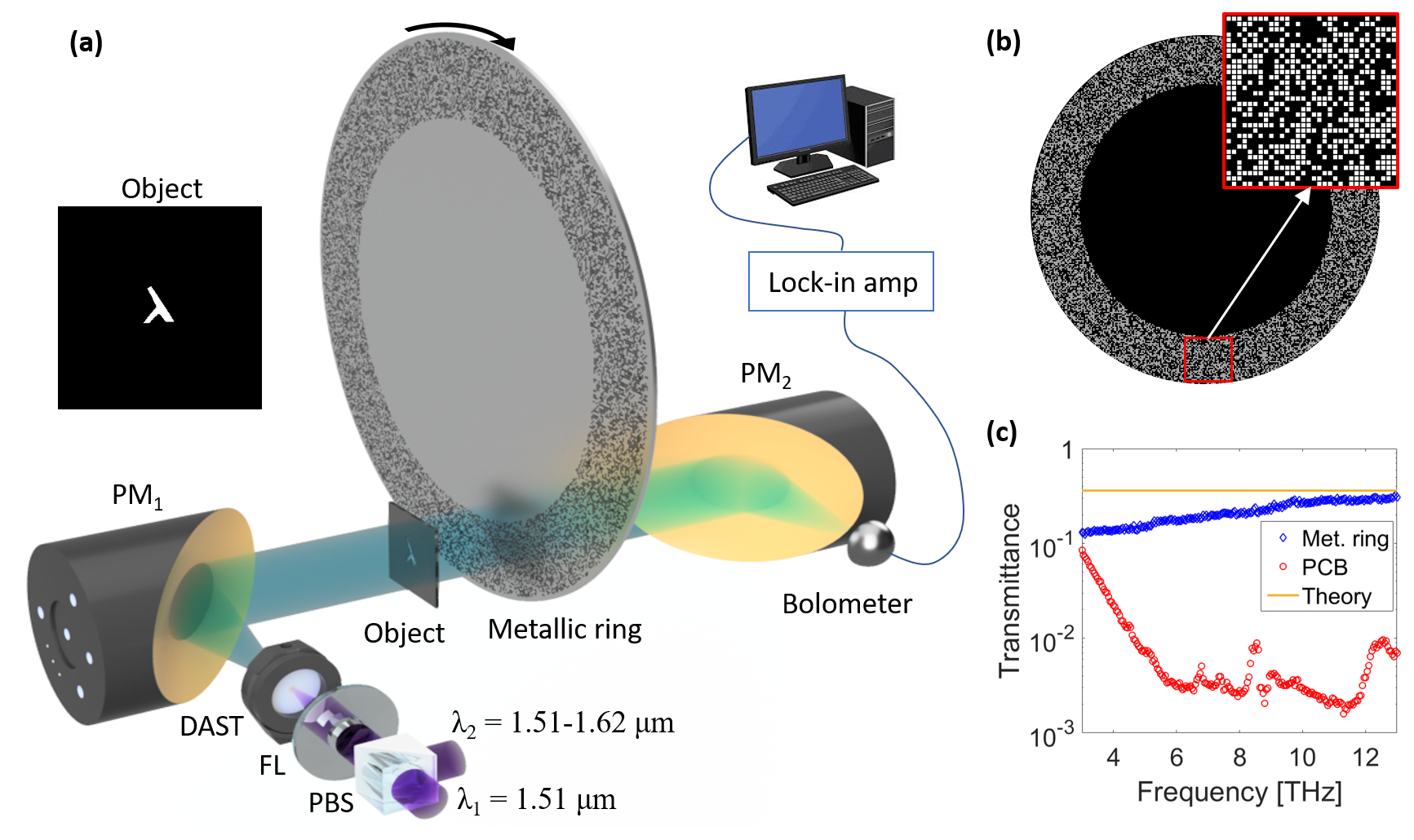}
\caption{(a) Conceptual drawing of the experimental setup used for SPI in the terahertz region. PBS: polarization beam splitter; $\lambda_1$ and $\lambda_2$ : input wavelengths; FL: Fourier lens; DAST: nonlinear crystal for difference frequency generation; PM$_1$ and PM$_2$ : parabolic mirrors ($f$ = 50 mm); Bolometer: single-pixel terahertz detector. Object's transmission profile shown in the inset. (b) Pattern encoded on the metallic ring and imaging area example (32 $\times$ 32 mm$^2$) in the inset. (c) Transmission of the imaging area within the metallic ring (blue diamonds) and of a standard PCB substrate (red circles). The orange horizontal line corresponds to the theoretical transmittance of the metallic ring, i.e., $0.5 \times 0.85^2 \approx 36\%$.}
\label{fig:setup}
\end{figure}

\subsection{Experimental setup}

Figure~\ref{fig:setup}(a) shows a schematic diagram of the experimental setup for testing our single-pixel imaging system. A monochromatic terahertz source \cite{miyamoto2008widely, miyamoto2019generation}, that comprises a 4-dimethylamino-N-methyl-4-stilbazolium tosylate (DAST) difference frequency generation (DFG) system was used (lasing frequency: 3-13 THz; pulse repetition frequency (PRF): 1 MHz; pulse width: 8.3 ps \cite{miyamoto2019generation}). Its output beam was collimated by a parabolic mirror (PM$_1$) and directed towards a partially opaque object and the metallic ring (photographs shown in the Appendix). The inset of Fig.~\ref{fig:setup}(a) shows the transmission intensity profile of the $\lambda$-shaped object obtained by saturating a CCD camera after illuminating the object with visible light. The distance between the object and the metallic ring was 20 mm.

The transmitted terahertz remaining signal was then focused by a second parabolic mirror (PM$_2$) onto a 4 K Si bolometer (Infrared Laboratories, Inc.), i.e., a broadband high-sensitivity terahertz single-point detector, so the beam was smaller than the bolometer's detecting area. It is worth noting that a lock-in amplifier was used to reduce signal output noise.



The metallic ring used in this experiment was formed from a 0.15-mm-thick stainless steel (SUS) plate directly perforated without any substrate holding the random masks. It was manufactured using a solder mask for a re-flow soldering process \cite{code} to avoid undesired absorption within the terahertz spectrum region of 3-13 THz. The 35-mm-wide ring included randomly encoded ON (white) and OFF (black) pixels with dimensions of 1 $\times$ 1 mm$^2$ and almost equal probabilities, as shown in Fig.~\ref{fig:setup}(b). The ON pixels were formed by an aperture (hole) with dimensions of 0.85 $\times$ 0.85 mm$^2$, formed by a 0.15-mm frame. Note that the ring was supported by a metallic disk without ON pixels to improve its robustness; however, such design is not essential and can be redesigned in future implementations being able to consider a wider ring. The red square shows an example of the imaging area used (32 $\times$ 32 mm$^2$) for a particular rotation.

The ring acted as a diffraction grating to attenuate the terahertz power by one-third for lower terahertz frequencies. This is because the frame thickness, i.e., 0.15 mm, is on the same order of magnitude as the wavelengths in the lower terahertz region. However, its loss was approximately two orders of magnitude less than that of previously reported standard PCB spinning disks \cite{chan2008single, shen2012spinning}, as shown in the transmission curves of Fig.~\ref{fig:setup}(c). The loss in our case can be considered to be almost constant $\mathcal{T}$ in Eq.~(\ref{eqn:randommaskimage}) for the entire terahertz frequency range. The transmission curves in Fig.~\ref{fig:setup}(c) were obtained by removing the object but keeping the metallic ring at a fixed angle (blue diamonds), or removing it and introducing a standard PCB substrate without any printed copper instead (red circles), while tuning the source from 3 to 13 THz. Each detected signal was then compared with a reference intensity curve considering no objects nor masks between the two parabolic mirrors.

The theoretical maximum transmission possible using our ring design is not $50\%$, but $0.5 \times 0.85^2 \approx 36\%$ (see orange line in Fig.~\ref{fig:setup}(c)). This is due to having $50\%$ of ON pixels with hole dimensions of 0.85 $\times$ 0.85 mm$^2$ out of a total pixel size of 1 $\times$ 1 mm$^2$. Note that the diffraction losses and theoretical maximum transmission can be improved by changing the pattern design and increasing the ON:OFF pixel ratio.

\begin{figure}[t!]
\centering
\includegraphics[width=\linewidth]{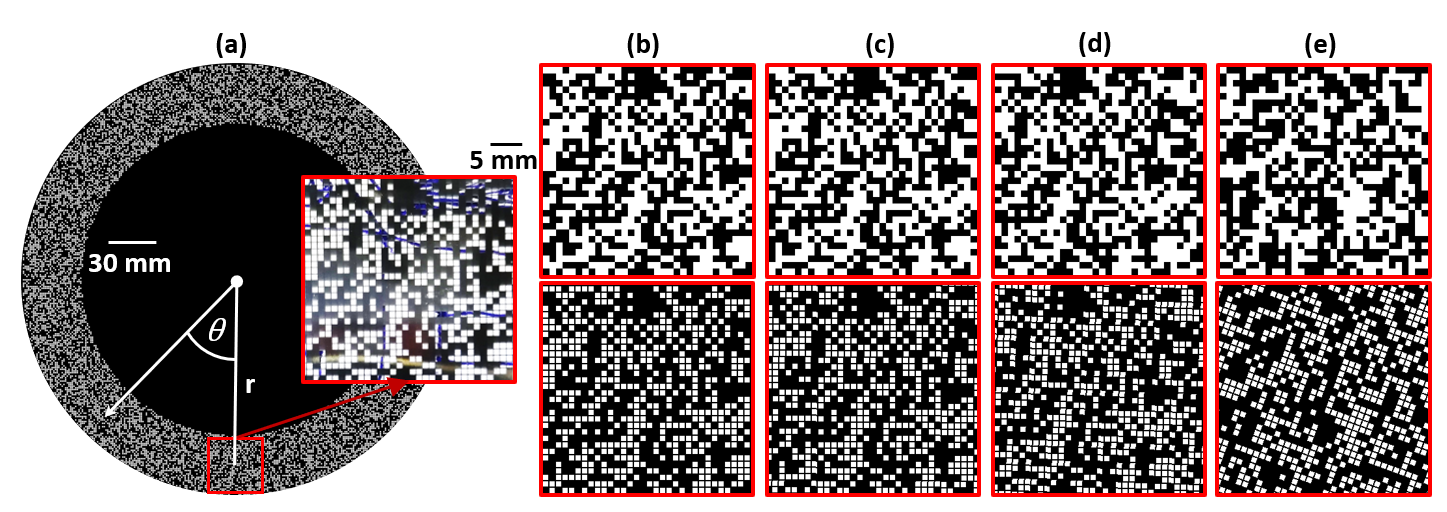}
\caption{Digitization of the imaging area from the \textit{physical} random masks. (a) Metallic ring (r = 112 mm) with a photograph of the \textit{physical} mask over a simple drawing shown in the inset, and \textit{digital} masks mimicking the imaging area for disk angles of (b) $\theta$ = 0$^\circ$, (c) 0.2$^\circ$, (d) 2$^\circ$, and (e) 20$^\circ$. Top and bottom rows show the low- (32 $\times$ 32) and high- (320 $\times$ 320) pixel-resolution masks, respectively.}
\label{fig:digit}
\end{figure}

Different random patterns are generated by rotating the whole ring while fixing the transverse position of the imaging area, as shown in the digitized metallic disk design of Fig.~\ref{fig:digit}. A photograph of the \textit{physical} mask transmission over a simple drawing is shown for comparison in the inset of Fig.~\ref{fig:digit}(a), and its digitization considering various angles ($\theta$) and pixel resolutions in Figs.~\ref{fig:digit}(b-e). The object can be then reconstructed by rotating the whole disk and displacing it vertically after each full rotation to slightly improve the image quality. An animation of a reconstruction example using experimental data is given in the Supplementary Material to help understand such SPI scheme operation (see \textcolor{urlblue}{Visualization 1}).

\section{Results}

\subsection{Higher pixel resolution due to subpixel digitization}

The number of pixels in the \textit{digital} random masks used in the object reconstruction algorithm, $\mathbf{\Phi}$ in Eq.~(\ref{eqn:randommaskimage}), determines the final image resolution. If we apply 32 $\times$ 32 pixel masks, like those in the top row of Figs.~\ref{fig:digit}(b-e) and typically used in previous terahertz SPI schemes \cite{chan2008single, shen2012spinning, shrekenhamer2013terahertz}, we can quickly reconstruct any object with only a few measurements (see extended analysis of pixel resolution and measurements needed in the Appendix).

The introduction of new random elements in $\mathbf{\Phi}$ is achieved by rotating the \textit{physical} metallic ring, namely the division of a \textit{physical} pixel into several \textit{digital} subpixels, allowing us to arbitrarily increase the pixel resolution of the \textit{digital} random masks. A different subpixel concept for SPI systems \cite{sun2016improving} effectively reduces the SNR by laterally displacing the encoded masks in a digital micromirror device (DMD). However, the pixel resolution in that case is limited by the finite number of pixels in the DMD screen. The bottom row of Figs.~\ref{fig:digit}(b-e) shows the high pixel resolution digitization example we use for most of the THz imaging results shon in this proof-of-concept experiment, considering the same angles as those in the top row, but subdividing each \textit{physical} pixel by 10 $\times$ 10, resulting in 320 $\times$ 320 pixel \textit{digital} masks $\mathbf{\Phi}$. Thus, we can approximately simulate the frames improving the image definition, but also increase the similarity between the real spinning disk pattern and its digitization for all angles, as clearly shown in Fig.~\ref{fig:digit}(e) for the $\theta$ = 20$^\circ$ case.

\begin{figure}[t!]
\centering
\includegraphics[width=\linewidth]{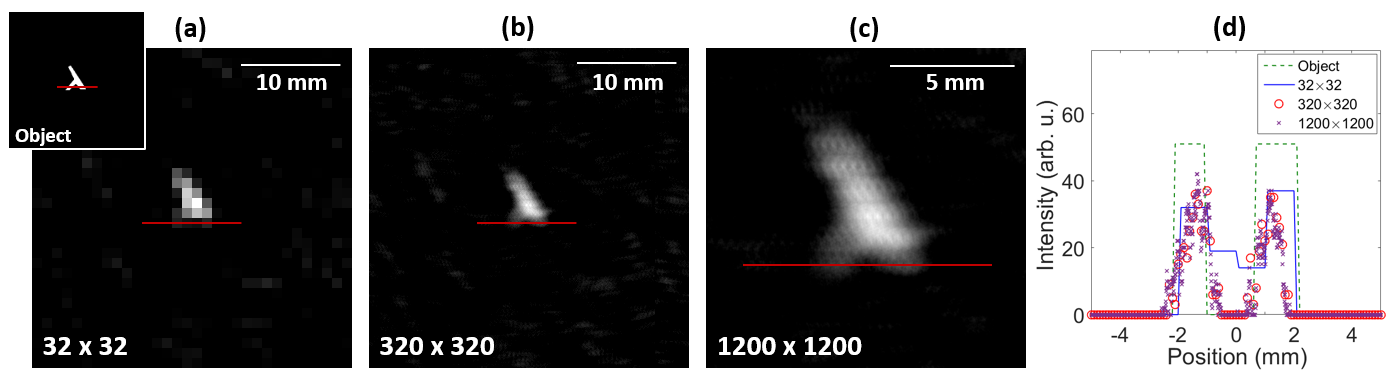}
\caption{Intensity profiles of the $\lambda$-shaped object (inset) and its experimental SPI reconstruction using 13 THz radiation when considering the (a) 32 $\times$ 32 and (b) 320 $\times$ 320 pixel \textit{digital} masks described in Fig. \ref{fig:digit}, and (c) considering a smaller imaging area (12 $\times$ 12 mm$^2$) but subdividing each \textit{physical} pixel by 100 $\times$ 100, obtaining a final pixel resolution of 1200 $\times$ 1200. Red horizontal lines in (a-c) show cross section for profiles plotted in (d). The object intensity in (d) has been attenuated 80\% to better approximate the overlap of the object in the inset of (a) with a Gaussian beam.}
\label{fig:profile}
\end{figure}

We show in the THz imaging experimental example of Fig.~\ref{fig:profile} how the image spatial details can be substantially improved by increasing the pixel resolution of $\mathbf{\Phi}$ in Eq.~(\ref{eqn:randommaskimage}), thanks to the subpixel digitization technique. Figures~\ref{fig:profile}(a) and (b) show the reconstructed images when illuminating the object (transmission intensity profile given in the inset) with 13 THz radiation and considering 32 $\times$ 32 or 320 $\times$ 320 pixel \textit{digital} masks $\mathbf{\Phi}$, respectively, as described in Fig.~\ref{fig:digit}. Even higher resolution image reconstruction is given in Fig.~\ref{fig:profile}(c) to show the full potential of this SPI system, considering a smaller imaging area (12 $\times$ 12 mm$^2$) but subdividing each \textit{physical} pixel by 100 $\times$ 100, obtaining a final pixel resolution of 1200 $\times$ 1200. We used the same number of masks in all image reconstructions for comparison, rotating the whole metallic ring every $\theta = 0.2 ^\circ$ and for 5 different vertical positions (total number of measurements $M$ = 9000, more information on the measurement mechanism in the following sections). Red horizontal lines show the cross section for the profiles plotted in Fig.~\ref{fig:profile}(d), being able to distinguish the difference in spatial resolution observed between low- (32 $\times$ 32) and high- (320 $\times$ 320) pixel-resolution cases.

The intensity profile of the reconstructed image when using high pixel resolution masks is considerably similar to the object's profile if considering the overlap with a Gaussian beam, having no intensity between the two peaks from the $\lambda$ 'legs'. Note that the horizontal and vertical darker lines in the reconstructed images, e.g. in Fig.~\ref{fig:profile}(c), are due to the frames and angle steps, respectively. Varying the step sizes could help removing these image imperfections.



\begin{figure}[t!]
\centering
\includegraphics[width=\linewidth]{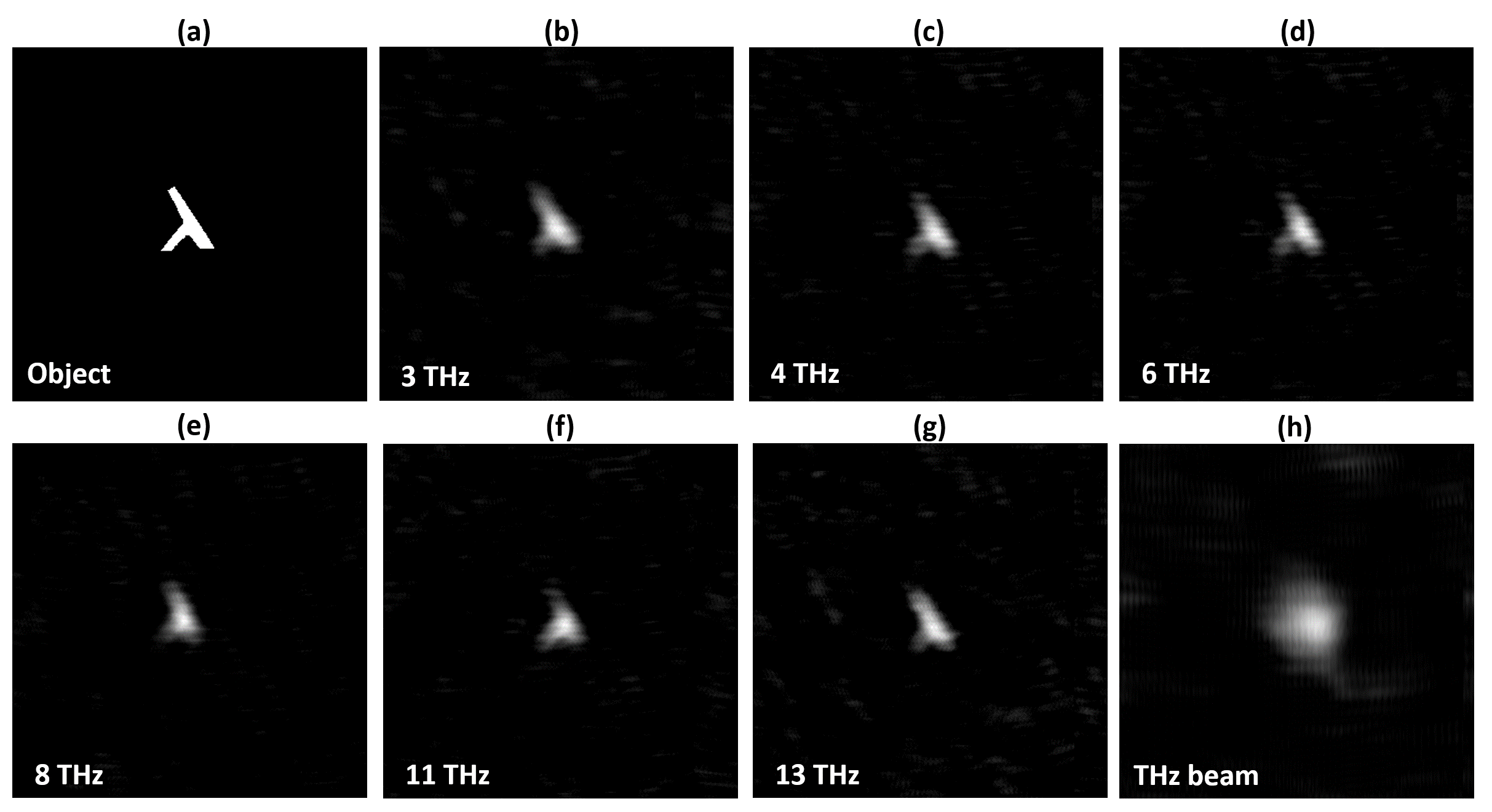}
\caption{Experimental results showing the broadband high-resolution imaging (320 $\times$ 320 pixels) capabilities using our SPI scheme for different bands within the high-frequency terahertz region (3-13 THz). (a) Intensity profile of the object used for the SPI reconstruction in the (b) 3, (c) 4, (d) 6, (e) 8, (f) 11, and (g) 13 THz regions. (h) Terahertz Gaussian beam for the 11 THz case.}
\label{fig:resLambda}
\end{figure}

\subsection{Broadband high-definition terahertz imaging results}

Figure~\ref{fig:resLambda}(a) shows the transmission intensity profile of the $\lambda$-shaped object, and its reconstructed images within a frequency region of 3-13 THz in Figs.~\ref{fig:resLambda}(b-g). This is the first demonstration, to the best of our knowledge, of a terahertz SPI system operating in the whole terahertz frequency region (3-13 THz). The reconstructed images were obtained by also rotating the metallic ring in increments of $\theta = 0.2 ^\circ$, being able to further increase the image quality by repeating the whole rotation for five vertical positions 1 mm apart. The total number of measurements was $M = 5\times1800 = 9000$; this value is still below the $N = 320 \times 320 \approx 10^5$ measurements ($M<N/10$) needed for a pixel-by-pixel raster scanning (RS) SPI scheme. The whole reconstruction process took around $T = 210$ minutes, mostly due to the slow rotation of the disk. The integration time was 400 ms and the time needed to complete each 0.2$^\circ$ rotation was approximately 1 s ($T = 1.4\times9000$ s). Note that the results presented here are a proof-of-concept showing the possibilities of this THz imaging system, and the total acquisition time could be easily decreased by optimizing the reconstruction algorithm (needing less measurements as described in the Appendix), and also by using a faster rotor being able to acquire the intensity data while rotating the whole metallic ring, avoiding the current "stop-measure-rotate" acquisition method used.

\begin{figure}[t!]
\centering
\includegraphics[width=\linewidth]{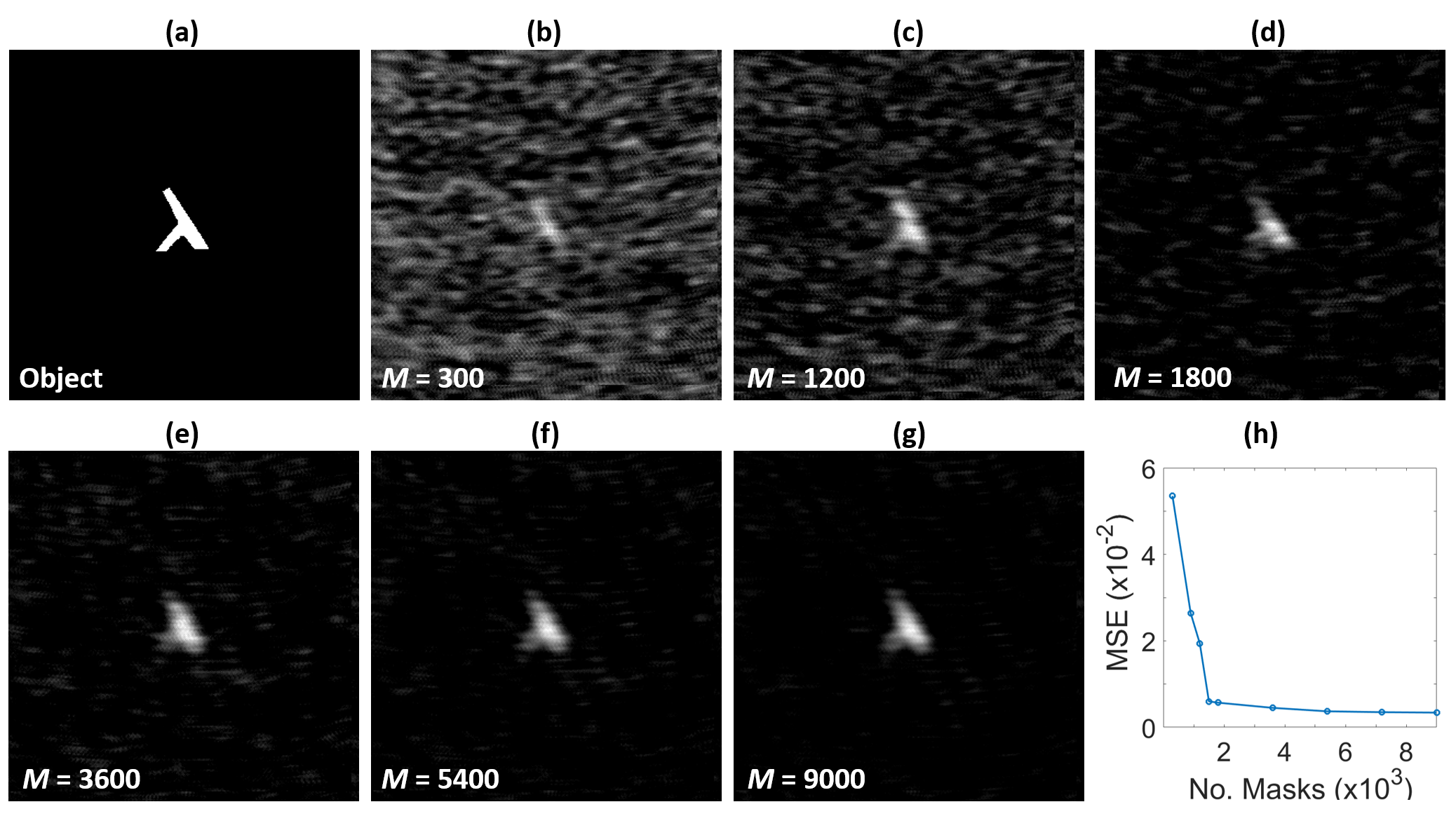}
\caption{Mean squared error (MSE) analysis for the 4 THz reconstruction results (see \textcolor{urlblue}{Visualization 1}). (a) 320 $\times$ 320 pixel object and its SPI reconstructed image using (b) $M$ = 300, (c) 1200, (d) 1800, (e) 3600, (f) 5400, and (g) 9000. (h) MSE values versus the number of masks used (MSE = 0.0535, 0.0194, 0.0057, 0.0045, 0.0037, and 0.0034 for 300, 1200, 1800, 3600, 5400, and 9000 masks, respectively).}
\label{fig:MSE}
\end{figure}

Terahertz output radiation at higher frequency produced a narrower collimated beam owing to diffraction effects, as shown in the example in Fig.~\ref{fig:resLambda}(h) for 11 THz. That is why we used a $\lambda$-shaped object, for which the most relevant information is concentrated in the center of the object (see the Appendix for an example of a larger object using IR laser light), thereby enabling the rapid recognition of the image shape with high pixel resolution. We show in Fig.~\ref{fig:MSE}, how the $\lambda$-shaped object can be recognized after only $M = 300$ measurements ($M \approx N/300$), corresponding to $T < 8$ minutes. 

There are several ways to quantify the reconstructed image quality or the level of noise with respect to the signal. Here, we consider the mean squared error (MSE) \cite{candes2006compressive} obtained from a pixel-by-pixel comparison of the object (see Fig.~\ref{fig:MSE}(a)) and its reconstructed image (see Figs.~\ref{fig:MSE}(b-g)) using Eq.~(\ref{eqn:randommaskimage}). The MSE decreases as the object ($\mathbf{O}^\prime$) and its reconstructed image ($\mathbf{O}$) converge
\begin{equation}
  \text{MSE} = \frac{\sum_{i,j}^{n} \: \left[O_{i,j}^\prime - O_{i,j}\right]^2}{\sum_{i,j}^{n} \: O_{i,j}^2},
\label{eqn:mse}
\end{equation}
where $O_{i,j}^\prime$ and $O_{i,j}$ are the object reference and its measured output of the reconstructed image at pixel ($i,j$), respectively, and $n$ is the number of pixels along the $x$ and $y$ axes.


As shown in Fig.~\ref{fig:MSE}, the MSE sharply decreases from $\sim0.06$ to $\sim0.006$ after the first revolution is completed in increments of $\theta$ = 0.2$^\circ$ ($M$ = 1800). A further increase in $M$ produced higher-quality images without any background noise, with minimum MSE = 0.0034 for $M$ = 9000. The slight MSE saturation after the first revolution ($M$ = 1800), shown in Fig.~\ref{fig:MSE}(h), suggests that longer vertical displacements would notably improve the convergence between $\mathbf{O}'$ and $\mathbf{O}$. Note that all the SPI results shown in Fig.~\ref{fig:resLambda} have MSE values of below 0.004.

\subsection{Spectral imaging}


\begin{figure}[t!]
\centering
\includegraphics[width=\linewidth]{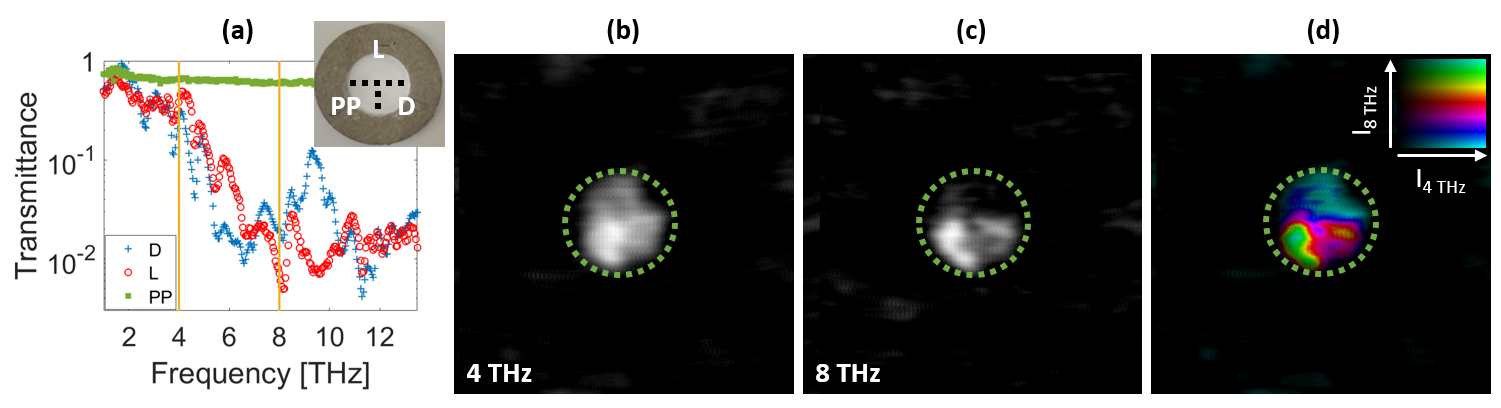}
\caption{Experimental results showing the broadband high pixel resolution imaging capabilities using our SPI scheme in the terahertz region for D-glucose, L-glucose, and polypropylene (PP). (a) Transmittance measured as in Fig.~\ref{fig:setup}(c) and photograph of the sample in the inset. 2D spectral imaging results obtained at (b) 4 and (c) 8 THz. (d) Molecule position identification by combining brightness from (b) and hue from (c).}
\label{fig:resSpect}
\end{figure}

Figure~\ref{fig:resSpect} shows an example of how we can perform a 2D spectral analysis of a particular sample. We used three materials, namely D-glucose, L-glucose, and polypropylene (PP), which look like white powder in the visible; however, they have different transmissions in the terahertz region, as shown in the measured transmission spectra of Fig.~\ref{fig:resSpect}(a). D-glucose, L-glucose, and PP were respectively purchased from Kanto Chemical Co. Inc. (CAS No. 50-99-7), Tokyo Chemical Industry Co., Ltd. (CAS No. 921-60-8), and Spectra Design Co., Ltd. (SD-PPW5). The powders of these chemical components were placed inside the bottom-right, top-most, and bottom-left sections of a plate with $\sim$1 mm aperture, respectively, as shown in the photograph of the inset in Fig.~\ref{fig:resSpect}(a). The plate was then pressed by a compression molding machine, obtaining a thickness of approximately 0.3 mm. A plate made from these molecules looks uniformly transparent after applying pressure, i.e., without any domains in the visible spectrum; however, PP exhibits high transmission in the frequency region of 3-13 THz, whereas D- and L-glucose show relatively low transmittance at 4 THz ($\sim$0.3 and $\sim$0.4, respectively), and even lower transmittance at 8 THz ($\sim$0.007 and $\sim$0.02, respectively), as shown in the two regions indicated by the orange vertical lines in Fig.~\ref{fig:resSpect}(a). Naturally existing glucose, i.e., D-glucose, contains mostly $\alpha$-phase molecules \cite{brauer2011vibrational}; however, its synthetically produced enantiomer, L-glucose, includes equal quantity of both $\alpha$- and $\beta$-phase molecules. Thus, 4 THz radiation can be used to distinguish PP from glucose, while 6-10 THz radiation (8 THz in our case), plays an important role in identifying D- and L-glucose. It should be noted that the selective synthetic control of the L-glucose form is difficult, furthermore, the spectra in the high frequency THz region is not available in the THz database \cite{web}. The high sensitivity and broadband capabilities of our THz spectroscopy system makes it a powerful tool to identify if the target material is natural or synthesized.

Figures~\ref{fig:resSpect}(b) and (c) show the reconstructed images at 4 and 8 THz, respectively, enabling the assignment of the exact allocation of the molecules. Although the boundary between PP and L-glucose is hard to recognize, our THz SPI system succeeded in the two-dimensional identification of molecules with the spatial resolution shown in Fig.~\ref{fig:profile}(d), allowing to distinguish shape inhomogeneities due to the different quantity of deposited powder when preparing the sample. We combined the normalized intensity by assigning the transmission images in Figs.~\ref{fig:resSpect}(b) and (c) to brightness and hue (rainbow colormap), respectively, obtaining the new image in (d) where the three molecules deposition sections are clearly distinguished. Note that the measurement frequencies were selected by considering not only the spectral difference between the two glucose, but also the spectral dependence of the output power that can be obtained from the DAST-DFG source \cite{miyamoto2014direct}.

\section{Discussion}

Even though, the highest pixel resolution used (1200 $\times$ 1200) show a slight improvement in the image quality, as can be seen in Fig.~\ref{fig:profile}, higher image definition is bounded by the diffraction limit inherent to the large wavelength of the radiation used when performing THz imaging. Other THz imaging systems have small imaging area dimensions, thus limiting the maximum achievable spatial resolution. However, this could be easily overcome in our SPI system by choosing a larger imaging area (considering a larger metallic ring design with wider perforated area), allowing for an object magnification while being able also to increase the pixel resolution.

Unfortunately, the short focal length parabolic mirrors limited the $\lambda$-shaped object size we could use (7-mm-high and 5-mm-wide), but we considered a bigger imaging area, i.e., 32 $\times$ 32 mm$^2$, in most of our THz imaging results (see Figs.~\ref{fig:resLambda} and \ref{fig:MSE}) for comparison with previous SPI systems. We show in Table~\ref{Table1} a summary of previously reported similar THz imaging system (see a comprehensive review in Ref.~\cite{edgar2019principles} for more SPI examples in other spectral regions).

\begin{table}[ht]
	\centering
	\caption{Summary of previous related THz imaging systems.}
	\begin{tabular}{|c|c|c|c|}
		\hline
		Operation principle & Spectral range & Pixel resolution & Imaging area \\ 
		\hline
		\hline
		This work & 3-13 THz$^a$ & 1200 $\times$ 1200 & 32 $\times$ 32 mm$^2$ \\
		\hline
		THz-TDS SPI \cite{olivieri2020hyperspectral} & 0.5-2.5 THz & 128 $\times$ 128 & 4 $\times$ 4 mm$^2$ \\ 
		\hline

		THz spectrographic SPI \cite{shen2009terahertz} & 0.5-2 THz & 20 $\times$ 20 & 40 $\times$ 40 mm$^2$ \\ 
		\hline
		Cu\&PCB SPI \cite{chan2008single} & 0.1-1 THz & 32 $\times$ 32 & 32 $\times$ 32 mm$^2$ \\ 
		\hline
		Cu\&PCB spinning disk SPI \cite{shen2012spinning} & 0.1-1 THz & 32 $\times$ 32 & 32 $\times$ 32 mm$^2$ \\ 
		\hline
		Spatial THz modulator SPI \cite{shrekenhamer2013terahertz} & 0.2-4.6 THz & 31 $\times$ 33 & 10.2 $\times$ 10.8 mm$^2$ \\ 
		\hline
		THz-PTDH RS \cite{petrov2016application} & 0.05-1.3 THz & 60 $\times$ 60 & 24 $\times$ 24 mm$^2$ \\ 
		\hline
		THz-TDS RS \cite{buron2015terahertz} & 0.25-1.2 THz & 250 $\times$ 250 & 100 $\times$ 100 mm$^2$ \\ 
		\hline
		THz camera \cite{oda2015microbolometer} & 0.3-4.3 THz & 640 $\times$ 480 & 15 $\times$ 11.3 mm$^2$ \\ 
		\hline
	\end{tabular}
	
	\hspace{-7cm}{\footnotesize\noindent$^a$ Demonstrated also for IR and visible up to 532 nm}\\
	\label{Table1}
\end{table}

It is also worth noting that we could not use the THz camera to obtain any of the imaging results shown in this work, even for the simplest example of the terahertz beam shown in Fig.~\ref{fig:resLambda}(h), due to the low sensitivity of such 2D array imaging device \cite{oda2015microbolometer}. To give a practical example for comparison, the minimum detectable power at 4 THz for the THz camera is $\sim$ 3.6 $\mu$W/cm$^2$ (pixel size 23.5 $\mu$m), but the output average power of our source at its maximum spectral configuration, i.e., 4 THz, is $\sim$ 1.3 $\mu$W/cm$^2$ if we consider a flat top beam of 10 mm diameter. Hence, not able to detect it without building a demagnifying 4$f$-system, to relay the image plane onto the THz camera, even for the simplest and optimum case: without considering the thermal radiation noise from the equipment itself and any object nor mask attenuating the THz signal.

Altogether, our imaging system allowed 2D transmission measurements on strongly absorbing samples, enabling us to achieve a dynamic range >30 dB (only limited by the source, 60 dB can be achieved with higher input power), which is comparable to the high-performance THz-TDS system using the plasma phenomena \cite{d2014ultra}. However, the overall efficiency, and most importantly, the spatial resolution that can be obtained in THz-TDS systems is degraded by the nonlinear processes involved \cite{warner1968spatial, andrews1970ir}.

In light of these results, the broadband THz SPI spectral system presented in this work has the potential to improve the imaging properties of previous THz spectroscopy systems, being able to visualize 2D spectral information with high definition images, to study also macro-molecules such as polymers. Although this work is focused in intensity-only imaging applications, other spinning disk designs can be further pursued for high-definition phase retrieval using THz radiation~\cite{chan2008terahertz, stantchev2020real}.

\section{Conclusion}

\noindent We demonstrated a terahertz single pixel imaging (SPI) technique that allows the high pixel resolution (1200 $\times$ 1200) reconstruction of an object within an entire terahertz frequency range (3-13 THz) by employing a perforated metallic ring encoded with a random mask. We can improve the pixel resolution of the reconstructed image simply by increasing the number of pixels encoded in the \textit{digital} masks used in the reconstruction algorithm, due to the subpixel digitization technique, while using the same \textit{physical} metallic ring. This technique can potentially be extended to develop ultra-broadband terahertz imaging systems that produce lower-noise and higher-resolution images compared to those obtained using commercial THz cameras. The imaging system itself has, in principle, no spectral limits for imaging if the correct materials for the ring are used. It is worth noting that the metallic ring, which includes holes, does not inherently constrain the frequency range of light sources, and thus this system enables the reconstruction of high-quality images in more spectral regions than those for previously proposed SPI schemes \cite{chan2008single, shen2012spinning, shrekenhamer2013terahertz}, even having demonstrated its operation considering visible (532 nm) and near-infrared (1510 nm) regions (see characterization results in the Appendix).

Although the proposed SPI scheme would benefit from further improvement, such as increasing the scanning velocity, it can already be applied to advanced technology, such as broadband terahertz spectroscopic imaging beyond current SPI schemes. Commercially available spatial light modulators (SLMs) and digital micromirror devices (DMDs) do not cover all spectral regions and exhibit high diffraction losses in long-wavelength regions. Furthermore, we can improve the pixel resolution for a rotating random mask configuration without decreasing the detected intensity. The reconstruction algorithm can use any pixel resolution in $n \times n$ format for masks $\mathbf{\Phi}$, in Eq.~(\ref{eqn:randommaskimage}), as long as it is correlated with the correct angular position of the \textit{physical} metallic ring.



Such directly perforated random mask allow the development of ultra broadband 2D SPI systems that have applications such as the detection of harmful gas or pedestrians in imaging systems for self-driving cars. The ability to use such simple and inexpensive spatial projection for the visible and infrared regions may allow 2D imaging applications for which reducing costs while maintaining high pixel resolution is an important commercial goal. The light traversing the metallic ring can be frequency-multiplexed and detected using fast and inexpensive point detectors. Real-time images could thus be obtained using any band of the optical spectrum, from near ultraviolet to terahertz.

\section*{Funding}

This study received funding from the Japan Society for the Promotion of Science (16H06507, 18H03884, 19K05299, 18K04967) and the Japan Science and Technology Agency, Core Research for Evolutional Science and Technology (JST-CREST).

\section*{Disclosures}

The authors declare no conflicts of interest related to this article.

\section*{Appendix}

\subsection*{Single-pixel camera reconstruction optimization}


There are many methods to generate patterns for masking an object (with ON and OFF pixels that transmit and block the optical signal, respectively), including the Hadamard \cite{pratt1969hadamard}, Fourier \cite{zhang2015single}, and Toeplitz \cite{bajwa2007toeplitz} methods. There are also several ways to optimize the reconstruction algorithm, by reducing the total number of measurements needed ($M<N$), e.g., by minimizing the $\ell_1$-norm \cite{donoho1989uncertainty} or the total variation and curvature in an image \cite{rudin1992nonlinear, goldluecke2011introducing}. With an optimized reconstruction algorithm, reconstructed 32 $\times$ 32 pixel images can be obtained with much fewer measurements ($M \sim 200$), as shown in previous work \cite{shen2012spinning, edgar2019principles}. Such algorithms for SPI optimization slightly increase computation times.



The generated masks should be orthogonal to each other and equally ON:OFF pixel weighted for faster reconstruction. However, pseudo-random pattern reconstruction methods, i.e., non-orthogonal patterns with a small number of new pixels per mask, have been proven to be as effective as those using fully random patterns \cite{fornasier2010theoretical, haupt2010toeplitz}. Taking advantage of this principle, we make use of a circular random mask ring, where the introduction of new randomly situated pixels in the imaging area is due to the rotation of the whole ring-shaped structure. We considered a wide ring (r = 112 mm in Fig. \ref{fig:digit}(a)) for this proof-of-concept experiment because a large r value leads to a large number of newly introduced elements per angle when rotating the whole disk while keeping the imaging area static.

\subsection*{Subpixel digitization analysis}

\begin{figure}[htpb]
\centering
\includegraphics[width=\linewidth]{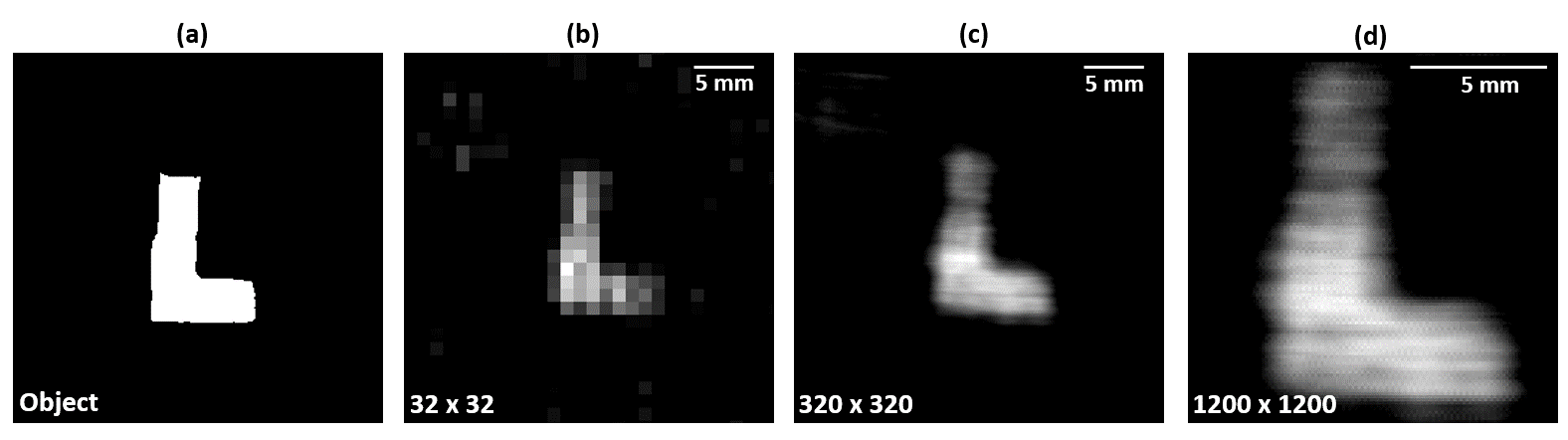}
\caption{Experimental examples of (a) 'L'-shaped object SPI reconstruction using a metallic ring and a 1510-nm laser source. (b) Low-pixel-resolution (32 $\times$ 32) image reconstruction using 900 measurements, (c) high pixel resolution (320 $\times$ 320) image reconstruction using 9000 measurements, and (d) very-high pixel resolution (1200 $\times$ 1200) image reconstruction using 9000 measurements (a smaller imaging area was used and \textit{pysical} pixels were subdivided by 100 $\times$ 100).}
\label{fig:resRes}
\end{figure}

The experimental examples in Fig.~\ref{fig:resRes} show the different reconstructed image resolutions that we can obtain by selecting the number of pixels in the digitization mask $\mathbf{\Phi}$ in Eq.~(\ref{eqn:randommaskimage}), considering a larger object. We use the same SPI system given in Fig.~\ref{fig:setup} but with an 'L'-shaped object and laser light at 1510 nm to illuminate the whole object (see Fig.~\ref{fig:resRes}(a) and photograph below).

If we consider lower-pixel-resolution \textit{digital} masks $\mathbf{\Phi}$, i.e., 32 $\times$ 32 pixel masks, we can quickly reconstruct any object with only a few steps. That is to say, we can consider that each \textit{physical} pixel corresponds to one \textit{digital} pixel, thus ignoring the frame holding the holes. In this case, we can reconstruct a 32 $\times$ 32 pixel image by rotating the metallic ring in increments of $\theta = 2^\circ$ for five vertical positions 1 mm apart ($M = 5 \times 180 = 900$ measurements). Figure~\ref{fig:resRes}(b) shows the reconstruction image of the object given in (a), for the 32 $\times$ 32 pixel masks case, as the ones in the examples in the top row of Figs.~\ref{fig:digit}(b-e).


We can increase the pixel resolution by subdividing each \textit{physical} pixel by 10 $\times$ 10 when encoding the \textit{digital} masks. With this composition, smaller rotations can be made while keeping fairly new random pixels in each new mask $\mathbf{\Phi}$. Figure~\ref{fig:resRes}(c) shows how we can reconstruct the same object with higher-pixel-resolution masks (see bottom row in Figs.~\ref{fig:digit}(b-e)). Similarly, we rotated the whole disk in increments of $\theta = 0.2^\circ$ for five different vertical positions, i.e., $M = 5 \times 1800 = 9000$ measurements, obtaining a 320 $\times$ 320 pixel resolution image. Note that the resolution can be changed to obtain even higher resolution images, which have the same detected signal on average. Figure~\ref{fig:resRes}(d) shows an experimental example of a 1200 $\times$ 1200 reconstructed image obtained using the same measurements as those in (c) but with a smaller imaging area (12 $\times$ 12 mm$^2$) and each \textit{physical} pixel subdivided by 100 $\times$ 100.

Although an experimental example of extremely high pixel resolution (1200 $\times$ 1200) is given to show the full capability of our scheme, \textit{digital} random masks with 320 $\times$ 320 pixels were used in most of the terahertz SPI results shown in this manuscript.

\subsection*{Masks and objects}

\begin{figure}[htpb]
\centering
\includegraphics[width=0.7\linewidth]{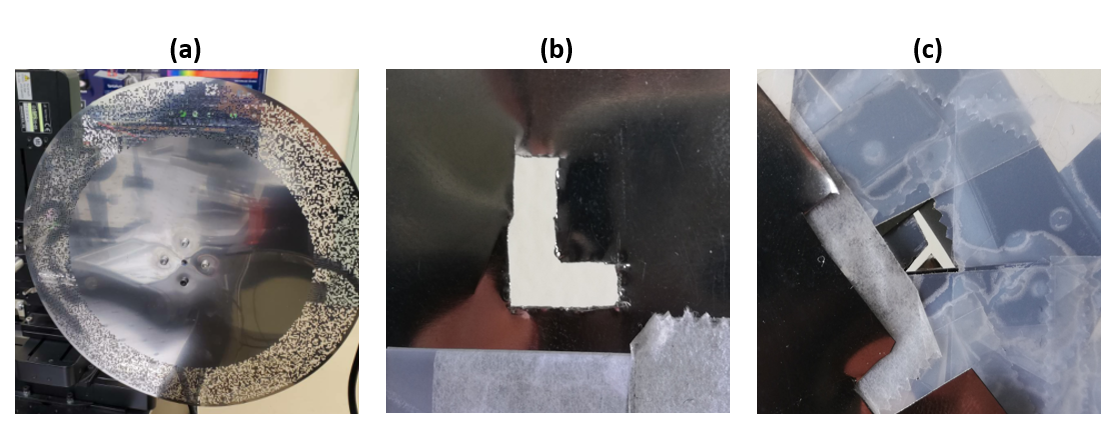}
\caption{Photographs of the (a) perforated metallic ring, (b) 'L'-shaped crafted and (c) $\lambda$-shaped assembled objects.}
\label{fig:Obj}
\end{figure}

Figure~\ref{fig:Obj} shows photographs of the metallic ring and 'L'- and $\lambda$-shaped objects. The hand-crafted 'L' object was too large to be used in the terahertz SPI experiments due to the small terahertz beam size (see Fig.~\ref{fig:resLambda}(h)). We thus decided to assemble a smaller $\lambda$-shaped object from some SUS leftovers. The diameter of the whole metallic ring was 256 mm. The 'L'-shaped object was 12-mm-high and 8.5-mm-wide and the $\lambda$-shaped object was 7-mm-high and 5-mm-wide.

\subsection*{Broadband SPI characterization}

\begin{figure}[hbtp]
\centering
\includegraphics[width=0.7\linewidth]{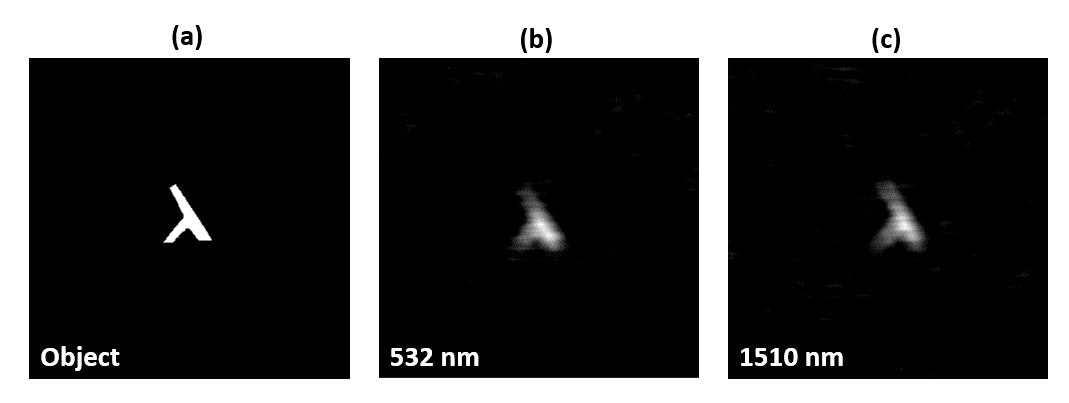}
\caption{Characterization measurements for the (a) $\lambda$-shaped object using (b) 532-nm and (c) 1510-nm laser light.}
\label{fig:charact}
\end{figure}

To characterize our broadband SPI scheme, we show in Fig.~\ref{fig:charact} the $\lambda$-shaped object reconstruction using laser light at 532 nm and 1510 nm. Here, we also rotated the metallic ring in increments of 0.2$^\circ$ for five vertical positions, as done for the reconstruction results in the main text. Note that when using visible and infrared light, we did not encounter the terahertz small beam size problem, and could thus distinguish the whole $\lambda$ shape.

\end{document}